\documentclass[prl,reprint,a4paper,superscriptaddress,citeautoscript,floatfix]{revtex4-1}

\pdfoutput=1
\usepackage[charter]{mathdesign}
\usepackage{amsmath}
\usepackage[pdftex]{graphicx}
\usepackage{microtype}
\usepackage{bm}\let\vec\bm

\usepackage[svgnames]{xcolor}
\usepackage[
	colorlinks=True,linkcolor=DarkRed,citecolor=ForestGreen,urlcolor=MediumBlue,
	pdfstartview=FitH,bookmarks=False,pdfpagemode=UseNone
]{hyperref}

\begin{document}

\title{\boldmath Observation of Caroli--de Gennes--Matricon Vortex States in YBa$_2$Cu$_3$O$_{7-\delta}$}

\author{Christophe Berthod}
\author{Ivan Maggio-Aprile}
\author{Jens Bru{\'e}r}
\affiliation{Department of Quantum Matter Physics, University of Geneva, 24 quai Ernest-Ansermet, 1211 Geneva, Switzerland}
\author{Andreas Erb}
\affiliation{Walther-Meissner-Institut, Bayerische Akademie der Wissenschaften, Walther-Meissner-Strasse 8, D-85748 Garching, Germany}
\author{Christoph Renner}
\affiliation{Department of Quantum Matter Physics, University of Geneva, 24 quai Ernest-Ansermet, 1211 Geneva, Switzerland}

\date{April 25, 2017}

\begin{abstract}

The copper oxides present the highest superconducting temperature and properties at odds with other compounds, suggestive of a fundamentally different superconductivity. In particular, the Abrikosov vortices fail to exhibit localized states expected and observed in all clean superconductors. We have explored the possibility that the elusive vortex-core signatures are actually present but weak. Combining local tunneling measurements with large-scale theoretical modeling, we positively identify the vortex states in YBa$_2$Cu$_3$O$_{7-\delta}$. We explain their spectrum and the observed variations thereof from one vortex to the next by considering the effects of nearby vortices and disorder in the vortex lattice. We argue that the superconductivity of copper oxides is conventional, but the spectroscopic signature does not look so because the superconducting carriers are a minority.

\end{abstract}

\maketitle

Type-II superconductors immersed in a magnetic field let quantized flux tubes  perforate them: the Abrikosov vortices. This remarkable property underlies and often limits many applications of superconductors. In 1964, Caroli, de Gennes, and Matricon used the Bardeen-Cooper-Schrieffer (BCS) theory of superconductivity to predict that vortices in type-II superconductors host a collection of localized electrons bound to their core \cite{Caroli-1964}. The direct observation of these localized states 25 years later by scanning tunneling spectroscopy (STS) is a spectacular verification of the BCS theory \cite{Hess-1989}. The formation of vortex-core bound states is an immediate consequence of the superconducting condensate being composed of electron pairs, while excitations in the vortex, being unpaired, have a different topology. Core states are, therefore, a robust property of superconductors, like edge states in topological insulators, irrespective of the origin and symmetry of the force that glues the electrons into pairs. In spectroscopy, they appear in the clean limit $\ell\gg\xi$ as a zero-bias peak in the local density of states (LDOS) at the vortex center, where $\ell$ and $\xi$ are the electron mean free path and superconducting coherence length, respectively, or as a structureless LDOS in the dirty limit $\ell\lesssim\xi$ \cite{Renner-1991}. Next to NbSe$_2$ \cite{Hess-1989, Suderow-2014}, the core states were seen by STS in several superconducting materials \cite{deWilde-1997b, Eskildsen-2002, Nishimori-2004, Guillamon-2008a, Zhou-2013, Du-2015}, including the pnictides, which are believed to host unconventional pairing \cite{Yin-2009, Shan-2011, Song-2011, Hanaguri-2012, Fan-2015}. The high-$T_c$ cuprates stand out as the only materials in which the vortex-core states have been looked for but not found. In YBa$_2$Cu$_3$O$_{7-\delta}$ (Y123), discrete finite-energy structures initially believed to be vortex states \cite{Maggio-Aprile-1995, Shibata-2003b, *Shibata-2010} were recently shown to be unrelated to vortices \cite{Bruer-2016}. In Bi$_2$Sr$_2$CaCu$_2$O$_{8+\delta}$ (Bi2212), the vortex cores present no trace of a robust zero-bias peak, but instead very weak finite-energy features apparently related to a charge-density wave order \cite{Renner-1998b, Hoogenboom-2000a, Pan-2000b, Matsuba-2003a, *Matsuba-2007, Levy-2005, Yoshizawa-2013}.

The absence of vortex states in cuprates is challenging the existing theories. Because these states are topological they are robust \cite{Wang-1995, Franz-1998b}, and they survive modifications of the BCS theory like strong-coupling extensions \cite{Berthod-2015} that do not change the nature of the condensate. To explain the cuprate vortex phenomenology, one needs either to leave BCS theory \cite{Arovas-1997, Himeda-1997, Andersen-2000, Kishine-2001, Berthod-2001b}, or to extend it by including additional order parameters that condense inside the vortex cores and gap out the zero-bias peak \cite{Zhu-2001a, Maska-2003, Fogelstrom-2011}. To date, none of these approaches has given a satisfactory account of the phenomenology observed by STS. The discovery that the low-energy structures in Y123 do not belong to vortices \cite{Bruer-2016} suggests that these theoretical efforts have been misguided.

The electronic structure of the cuprate high-$T_c$ superconductors (HTS) is notoriously complex, as manifested in a rich phase diagram. This complexity reveals a competition of different effective interactions, from which a variety of individual and collective modes emerges progressively as the temperature is lowered towards the ground state in which the system freezes at absolute zero, and which continues to keep the secret of the most stable superconductivity ever observed. It is generally believed that the phenomena taking place close to the Fermi surface---charge, spin, pairing orders, and their fluctuations---all derive from a single band \cite{Zhang-1988} or a small subset of bands \cite{Emery-1987} in the CuO$_2$ layer(s). Consistently, the interpretations of STS spectra \cite{Fischer-2007} have postulated that all electrons contributing to the measured LDOS are excited out of the superconducting condensate, in agreement with Leggett's theorem \cite{Leggett-1998}. On the other hand, it is well known that for all cuprates, at any doping, the superfluid density remains much smaller than the electron density \cite{Uemura-1989, Bernhard-1995, Bosovic-2016}. Our recent high-resolution STS experiments on Y123 have also revealed that only a fraction of the signal recorded on the sample surface is of superconducting origin \cite{Bruer-2016}. Early specific-heat measurements have given a similar hint \cite{Junod-1999}. We are therefore lead to a new paradigm, in which the low-energy electronic state of the HTS involves a minority superconducting channel in parallel with nonsuperconducting majority charges guilty for the pseudogap and the associated orders. Here we show that the minority carriers are fairly conventional in the superconducting state, showing Caroli--de Gennes--Matricon states in the vortex cores as predicted by the BCS theory for $d$-wave superconductors.

\begin{figure}[tb]
\includegraphics[width=0.75\columnwidth]{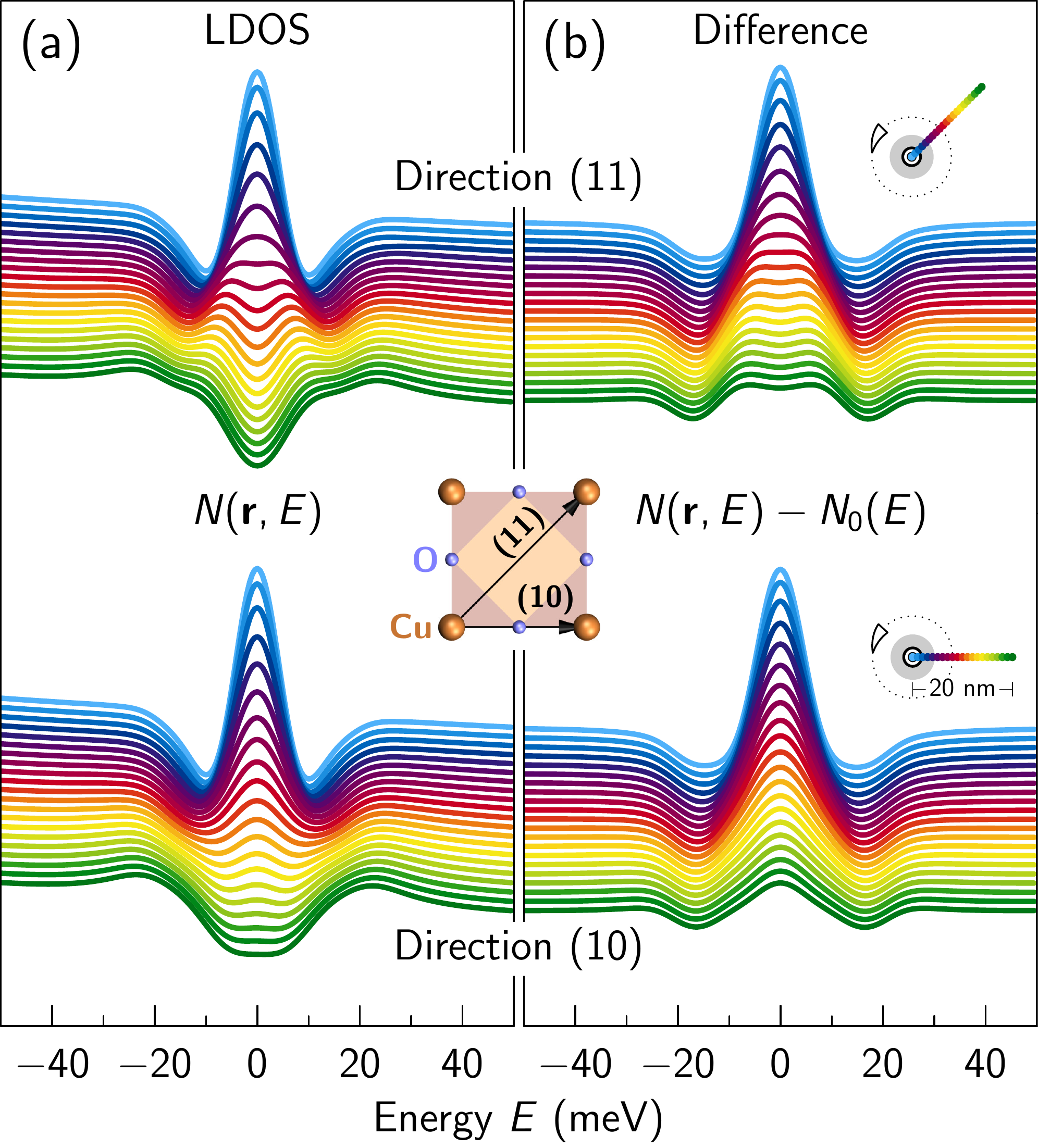}
\caption{\label{fig:fig1}
(a) The self-consistent LDOS calculated in the BCS theory for an isolated vortex in a $d$-wave superconductor with Y123 band structure is shown along two paths starting at the vortex core (blue) and ending at two points 20~nm away from the core (green) along the nodal (11) and antinodal (10) directions. (b) Same data with the zero-field DOS subtracted from all curves. The vertical scale is arbitrary and the curves are shifted vertically for clarity. The insets show the CuO$_2$ unit cell with the two crystallographic directions, and representations of the vortex core (gray disks and arrows indicating the supercurrent direction) with the color-coded paths of the two spectral traces.
}
\end{figure}

\begin{figure}[tb]
\includegraphics[width=\columnwidth]{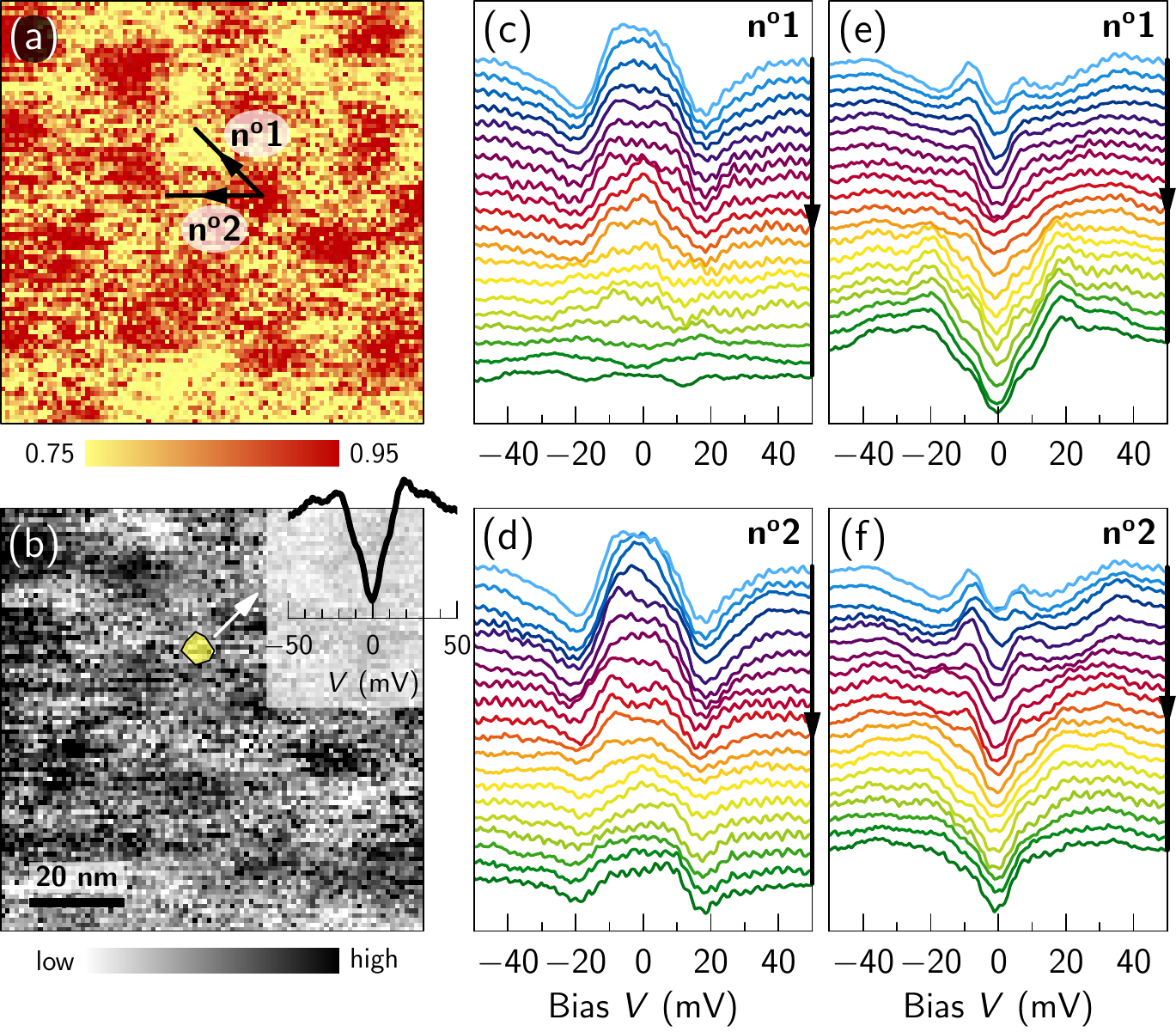}
\caption{\label{fig:fig2}
(a) and (b) A $90\times90~\mathrm{nm}^2$ area on the (001) surface of Y123 in a 6~T field, colored by (a) the ratio of the STS tunneling conductance at $+5$ and $+17$~mV, and (b) the conductance at zero bias. The inset in (b) shows the STS spectrum averaged over the small outlined region between vortices. Two series of difference spectra (raw STS data minus average spectrum shown in the inset) along the 20~nm paths indicated in (a) are displayed in (c) and (d). The color encodes distance from the core as in Fig.~\ref{fig:fig1}. Panels (e) and (f) show the raw $dI/dV$ data along the two paths.
}
\end{figure}

STS measurements with a normal metal tip do not discriminate the superconducting (SC) and nonsuperconducting (NSC) channels and collect electrons from each. Our working hypothesis is that the tunneling conductances originating from the SC and NSC channels are additive. An earlier report in which vortex-induced changes were tracked in Bi2212 rested on a different scenario, namely, the vortices would suppress locally the condensate and reveal a competing magnetic order \cite{Hoffman-2002}. As the NSC is not known, a simple subtraction to reveal the SC is not feasible. Yet, inhomogeneities of the SC like those induced by vortices can be singled out by subtracting the tunneling conductance outside vortices from that inside. If the formation of a vortex leaves the NSC unchanged, this procedure eliminates the NSC and permits a comparison of the SC inhomogeneities with the BCS theory. Figure~\ref{fig:fig1}(a) shows the LDOS predicted by the BCS theory at and near the center of an isolated vortex in a two-dimensional superconductor with electronic structure similar to that of Y123 \footnote{See Supplemental Material [url] for a presentation of the model and calculation methods, which includes 
Refs.~\onlinecite{Jorgensen-1990, Atkinson-2009, Schabel-1998a, *Schabel-1998b, Ichioka-1996, Torquato-2003, Torquato-2003, Weismann-2009}.}. The prominent feature is the zero-energy peak localized at the core center. With increasing distance from the core, the peak is suppressed with or without a splitting depending on the direction. Note that this LDOS anisotropy is unrelated to the $d$-wave gap anisotropy. In the quantum regime $k_{\mathrm{F}}\xi\sim1$ relevant for Y123, the vortex size is comparable with the Fermi wavelength and the Fermi-surface anisotropy determines the vortex structure \cite{Uranga-2016, Berthod-2016}. In the Supplementary Material, Fig.~\ref{fig:swave} indeed shows that the zero-bias LDOS is locked to the crystal rather than gap-node directions. For a meaningful comparison with experiment, the LDOS far from the vortex must be subtracted, as done in Fig.~\ref{fig:fig1}(b).

Figure~\ref{fig:fig2}(a) shows a $90\times90~\mathrm{nm}^2$ area on the surface of Y123, where 19 inhomogeneities can be identified as vortex cores. Details about the sample preparation and measurement technique were reported in Ref.~\onlinecite{Bruer-2016}. Because of the large NSC, the contrast due to vortices is weak; it is maximized by mapping the ratio of the STS tunneling conductance at 5 and 17 mV bias [Fig.~\ref{fig:fig2}(a)]. Local increases of the zero-bias conductance are also seen in the raw data [Fig.~\ref{fig:fig2}(b)], and correlate well with the vortex positions determined by the best contrast. In order to remove the NSC, we delineate a small region in-between vortices, calculate the average spectrum in this region [inset of Fig.~\ref{fig:fig2}(b)], and subtract this average from all spectra in the map. After subtraction, the spectral traces show the expected vortex signature with a maximum at zero bias in the cores. This is demonstrated in Figs.~\ref{fig:fig2}(c) and \ref{fig:fig2}(d) with two traces running from one vortex core along the two directions shown in Fig.~\ref{fig:fig2}(a). The peak developing locally at zero bias after subtracting the same background from the spectra of each trace clearly shows there is a larger local density at low energy near the vortex cores. It is not an artifact of the subtraction. Similar results are found in all vortices.

\begin{figure*}[tb]
\noindent\includegraphics[width=0.7\textwidth]{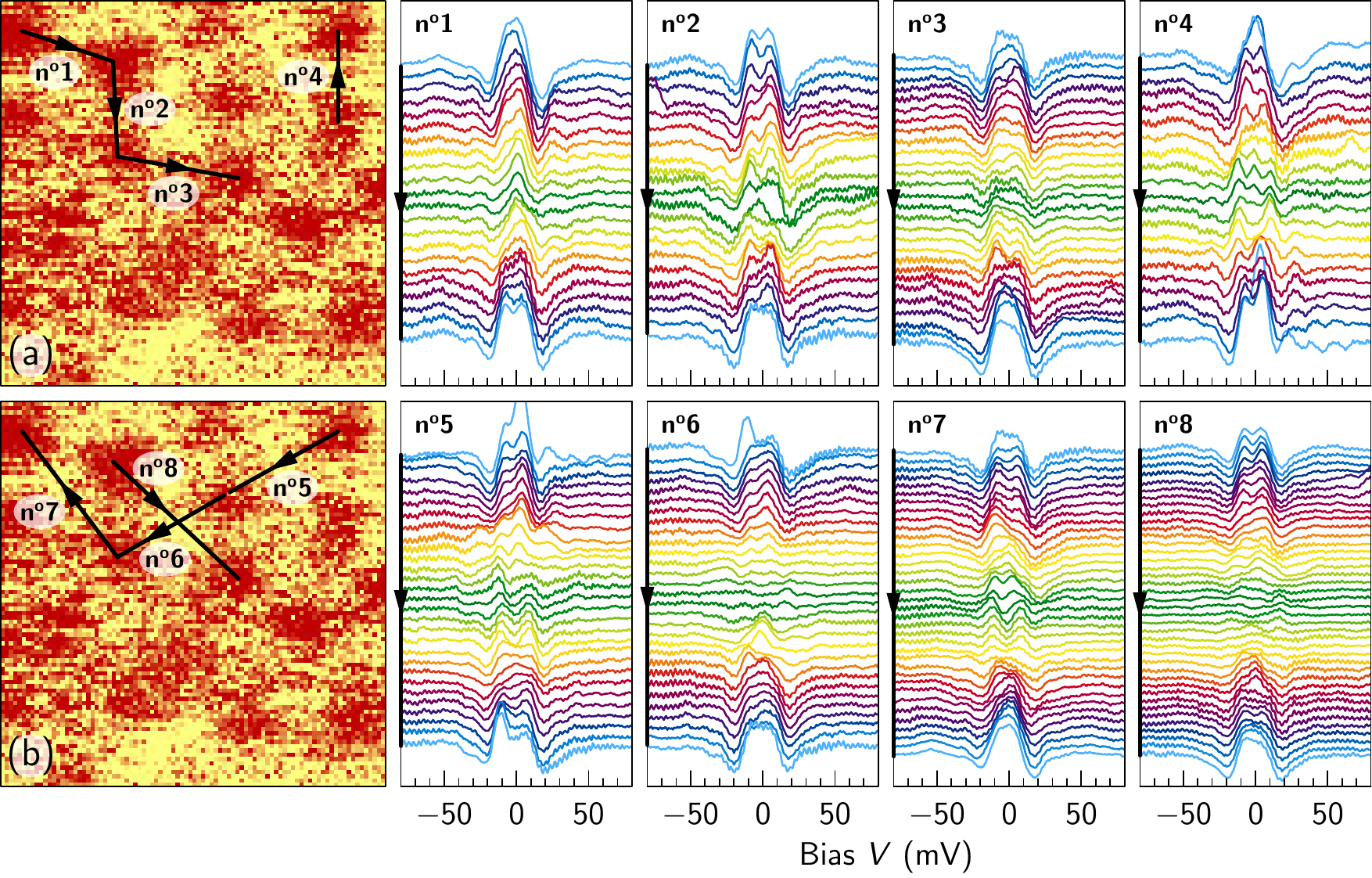}
\hfill\parbox[b]{0.25\textwidth}{
\caption{\label{fig:fig3}
Series of difference spectra [generated as in Figs.~\ref{fig:fig2}(c) and \ref{fig:fig2}(d)] along several paths connecting (a) nearest-neighbor and (b) next-nearest-neighbor vortices. The spectra are shifted vertically and colored from blue (vortex center) to green (in-between two vortices).
}
\vspace{4.75cm}}
\end{figure*}

A comparison of Figs.~\ref{fig:fig2}(c) and \ref{fig:fig2}(d) with Fig.~\ref{fig:fig1}(b) reveals evident similarities, but also differences. Similarities include the zero-bias peak, which has its maximum at the core center and is suppressed with increasing distance from the center, the absence of superconducting coherence peaks in the core leading to symmetric dips at $\pm17$~meV in both experiment and theory, a striking spatial anisotropy, with the peak extending farther along path n\textsuperscript{o}2 than along path n\textsuperscript{o}1, reminiscent of the different decay lengths observed in the theory between directions 10 and 11. Among the differences, one notices the central peak being taller in Fig.~\ref{fig:fig1}(b) than in the measurements, and the spectrum appearing locally reinforced or split at intermediate distances along path n\textsuperscript{o}1 and n\textsuperscript{o}2, respectively, while Fig.~\ref{fig:fig1}(b) shows a monotonic evolution. We now demonstrate that these differences can be explained by considering that (i) the vortices are not isolated in the experiment, (ii) the relative orientation of the vortex and crystal lattices influences the LDOS, and (iii) the vortex lattice is disordered. Incorporating (i) in the theory reduces drastically the calculated peak height; (ii) means that the LDOS anisotropy depends upon the positions of nearby vortices; finally, (iii) implies that each vortex sits in a specific local environment and presents spectra different from its neighbors.

Figure~\ref{fig:fig3} displays a series of spectral traces along various paths connecting either nearest-neighbor [Fig.~\ref{fig:fig3}(a)] or next-nearest neighbor [Fig.~\ref{fig:fig3}(b)] vortices. The notion of nearest- and next-nearest neighbor refers to a local fourfold coordination generally observed among the vortices, despite the long-range disorder in their arrangement. A trend is systematically observed: along a line connecting nearest-neighbor vortices, the zero-bias peak remains visible along the whole path, while it disappears along paths joining next-nearest neighbors. Considering the peak anisotropy as it is predicted by theory (Fig.~\ref{fig:fig1}), this trend suggests that the locally fourfold-coordinated vortex lattice tends to align along the crystalline axes, as we will confirm by a detailed modeling. In the absence of atomic resolution imaging, we infer the lattice orientation based on optical images of twin boundaries. Another lesson of Fig.~\ref{fig:fig3} is that all vortices, although similar, are different: some show a single peak, others show a split peak; the height of these peaks is also varying. This variability reflects disorder in the vortex positions, resulting in irregular distributions of supercurrents around each core. 

\begin{figure*}[tb]
\noindent\includegraphics[width=0.7\textwidth]{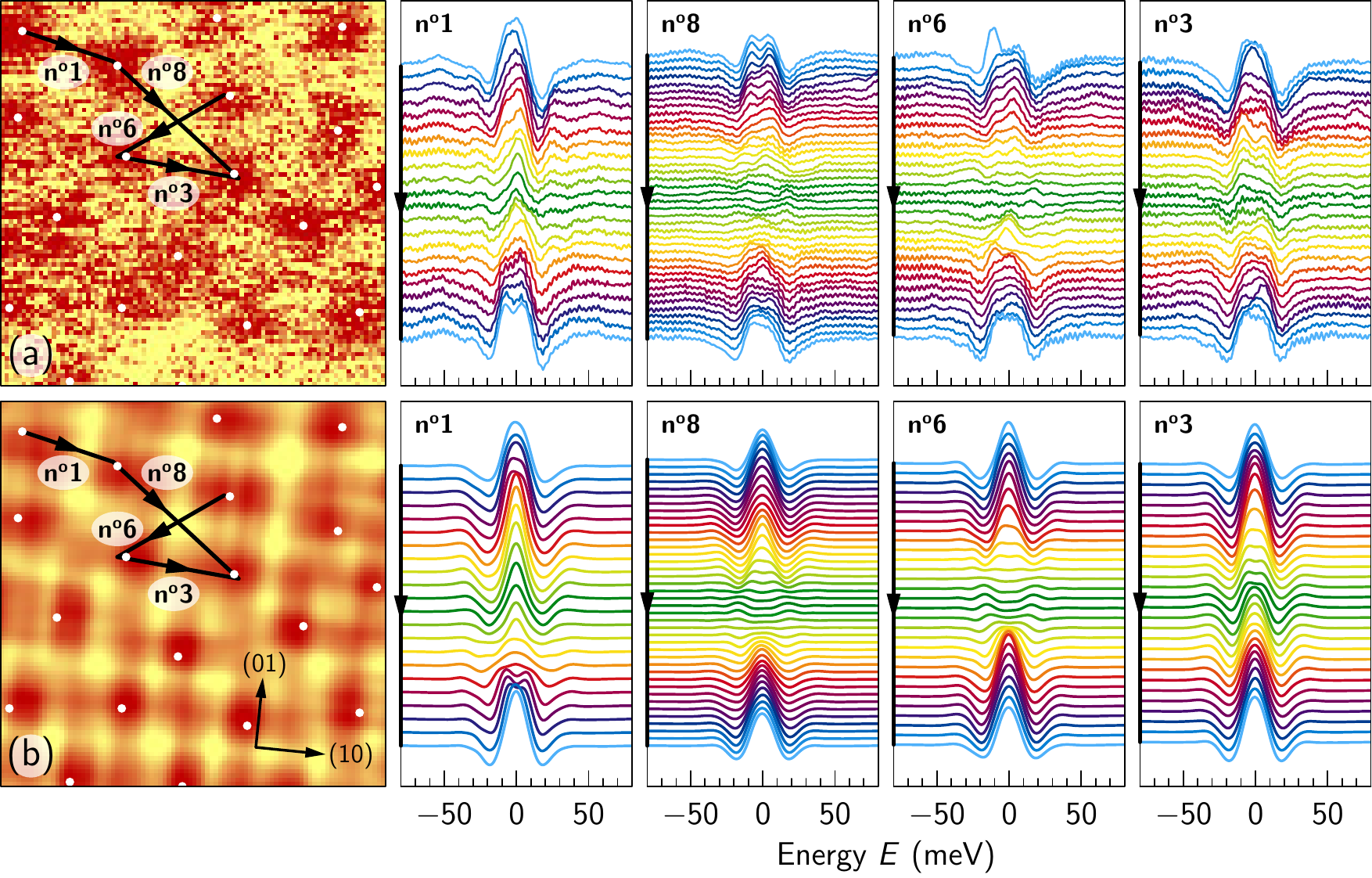}
\hfill\parbox[b]{0.25\textwidth}{
\caption{\label{fig:fig4}
(a) Same data as in Fig.~\ref{fig:fig3} for traces joining nearest-neighbor (n\textsuperscript{o}1 and n\textsuperscript{o}3) and next-nearest-neighbor vortices (n\textsuperscript{o}6 and n\textsuperscript{o}8). (b) Calculated ratio $[N(\vec{r},5~\mathrm{meV})+N_{\mathrm{NSC}}(5~\mathrm{meV})]/[N(\vec{r},17~\mathrm{meV})+N_{\mathrm{NSC}}(17~\mathrm{meV})]$, where $N(\vec{r},E)$ is the theoretical LDOS and $N_{\mathrm{NSC}}(E)$ represents the nonsuperconducting background \cite{Note1}. The simulated spectral traces are deduced from the theoretical LDOS by following the exact same procedure as applied to the experimental data. The white dots in (a) and (b) show the vortex positions, determined as the locations of maxima in (a) and corresponding to the phase singularity points of the order parameter in the simulation (b).
}
\vspace{0.3cm}}
\end{figure*}

We have undertaken large-scale simulations of the LDOS in disordered vortex configurations, in order to study how this modifies the core spectra with respect to the isolated vortex, and thus better understand the observations made by STS. Thanks to a new approach described in Ref.~\onlinecite{Berthod-2016}, we are able to compute the LDOS in a disordered vortex lattice with the same accuracy as in the isolated core. All relevant details are given as Supplemental Material \cite{Note1}, and we only briefly review here the key ingredients. The structure of the vortices in a finite field is deduced from the self-consistent solution in the ideal vortex lattice. We find that the LDOS changes dramatically with varying the orientation of the vortex lattice relative to the crystal axes. The system size required in order to compute the LDOS with our target resolution (3~meV) contains half a million unit cells and extends well beyond the area of our STS experiments, which contains roughly 54\,500 Cu sites. Inside the STS field of view, we locate vortices at the positions indicated in Fig.~\ref{fig:fig4}(a). Outside, we generate vortex positions randomly, however, constraining the intervortex distance to be at least 16~nm. The resulting configurations show no orientational order, but short-range coordination similar to what is seen in the STS image. We do not imply that the actual vortex distribution outside the field of view has no orientational order---in fact, it probably has some \cite{Maggio-Aprile-1995}---but the available data prevent us from inferring such an order. We also vary the orientation of the crystal lattice with respect to the vortices. For each of 600 generated configurations, we calculate the LDOS along the paths n\textsuperscript{o}6 and n\textsuperscript{o}8 of Fig.~\ref{fig:fig3} and compare with the experimental traces. With the configuration giving the smallest difference with experiment on these two traces, we recalculate the LDOS on the whole domain covered by STS with a resolution of 1~nm and deduce the theoretical map and traces shown Fig.~\ref{fig:fig4}(b).

The simulation confirms that quasiparticle scattering off nearby vortices reduces the central peak in each vortex, that the LDOS is different in all cores, and that it depends on the configuration of the vortices outside the STS field of view. We also find that the agreement with experiment is systematically better if the orientation of the microscopic lattice is such that the traces n\textsuperscript{o}6 and n\textsuperscript{o}8 are close to a nodal direction, in agreement with the twin-boundary directions. Although our search for a good vortex configuration has focused on two traces connecting next-nearest neighbor vortices, the resulting model reproduces the difference between these traces and those connecting nearest-neighbor vortices. It is also striking that the model correctly predicts the contrast of the STS image and the apparent size of the vortex cores without any further adjustments. Figure~\ref{fig:fig4}(b) presents less spectral variations from core to core than Fig.~\ref{fig:fig4}(a), but this appears to be a compromise by which this configuration achieves the best overall agreement. Other vortex configurations that compete closely do show variations comparable with experiment, including split peaks in some of the vortices (Fig.~\ref{fig:statistics}). We emphasize that our procedure does not deliver the best fit, which would require us to optimize systematically the vortex positions rather than trying random configurations. The vortex positions \emph{within} the field of view should be optimized as well. Figure~\ref{fig:fig4}(b) indeed reveals that the LDOS maxima are in general displaced with respect to the points where the order parameter vanishes: the LDOS gets polarized by asymmetries in the supercurrents \cite{Berthod-2013a, Berthod-2016}. Without any doubt, vortex configurations could be found that improve the agreement in Fig.~\ref{fig:fig4}, but we feel that such a costly optimization is unlikely to reveal new physics. Interestingly, the simulation presents vortex cores that appear to be split, e.g., at the beginning of path n\textsuperscript{o}6 (see also Fig.~\ref{fig:swave}). Reference~\onlinecite{Hoogenboom-2000b} reported a similar observation in Bi2212, which was ascribed to interaction with pinning centers. In Fig.~\ref{fig:fig4}(b) the splitting is merely a LDOS distortion due to an irregular distribution of the supercurrent.

In a clean BCS superconductor, the zero-temperature superfluid density $n_{s0}$ equals the total electron density $n$ \cite{Leggett-1998}. Estimates based on the penetration depth show instead that $n_{s0}/n=23\%$--$34\%$ in Y123 \footnote{At an optimal hole doping of 0.16, the nominal carrier density of the bilayer Y123 is $n=2\times0.84/(abc)=9.7~\mathrm{nm}^{-3}$ with the unit-cell parameters $(a,b,c)=(3.82,3.89,11.68)$~\AA. With an effective mass of typically 2--3 times the electron mass \cite{Padilla-2005} and a zero-temperature penetration depth $\lambda=1600$~\AA\ \cite{Basov-1995}, the estimated superfluid density $n_s=m^*/(\mu_0e^2\lambda^2)$ is 2.2--3.3~nm$^{-3}$.}. The analysis of STS data yields similar relations between the SC and NSC, with some model dependence, 14\% in Ref.~\onlinecite{Bruer-2016}, 25\% in the present work \cite{Note1}. In a one-channel picture, the property $n_{s0}\ll n$ requires large pairing fluctuations that would invalidate a mean-field description. Our results show that the mean-field theory works well in vortices, though. Another possibility would be that superconductivity emerges in a population of low-energy quasiparticles carrying only a small fraction of the spectral weight: in the RVB theory, the quasiparticle weight behaves as $2x/(1+x)$ as a function of doping $x$ \cite{Anderson-2004}, as observed in optical data of underdoped cuprates \cite{Mirzaei-2013}, which yields a value close to 25\% at optimal doping. In this scenario, the NSC should disappear with overdoping and the vortices should present a clear zero-bias peak, a challenge for future STS experiments. Alternatively, one can imagine two-channel scenarios. It has been recently proposed that the cuprate superconductivity is lead by oxygen $2p$ rather than copper $3d$ holes \cite{Rybicki-2016}. We speculate that the copper holes remain localized and form the NSC (in the specific case of Y123, where the zero-bias conductance is large, we cannot rule out a parasitic surface channel as another contribution to the NSC), while the oxygen holes are responsible for the recently uncovered universal Fermi-liquid signatures \cite{Barisic-2013} and enter the SC condensate. While the origin of pairing in this condensate remains mysterious, its spectroscopic properties are well described by the BCS theory, as we have demonstrated by unveiling the Caroli--de Gennes--Matricon-like states in the vortex cores.

We thank T.\ Giamarchi and D.\ van der Marel for their comments on the manuscript, N.\ Hussey and G.\ Deutscher for discussions, and G.\ Manfrini and A.\ Guipet for their technical assistance. This research was supported by the Swiss National Science Foundation under Division II. The calculations were performed in the University of Geneva with the clusters Mafalda and Baobab.

\let\oldaddcontentsline\addcontentsline%
\renewcommand{\addcontentsline}[3]{}%

\let\addcontentsline\oldaddcontentsline%

\onecolumngrid
\newpage
\begin{center}

{\large\textbf{\boldmath
Supplemental Material\\ [0.5em] {\small for} \\ [0.5em]
Observation of Caroli--de Gennes--Matricon Vortex States in YBa$_2$Cu$_3$O$_{7-\delta}$
}}\\[1.5em]

Christophe Berthod,$^1$ Ivan Maggio-Aprile,$^1$ Jens Bru{\'e}r,$^1$ Andreas Erb,$^2$ and Christoph Renner$^1$\\[0.5em]

\textit{\small
$^1$Department of Quantum Matter Physics, University of Geneva, 24 quai Ernest-Ansermet, 1211 Geneva, Switzerland\\
$^2$Walther-Meissner-Institut, Bayerische Akademie der Wissenschaften, Walther-Meissner-Strasse 8, D-85748 Garching, Germany
}

\vspace{2em}
\end{center}

\twocolumngrid
\renewcommand{\thefigure}{S\arabic{figure}}
\setcounter{figure}{0}

\tableofcontents

\section{\boldmath Microscopic model for the superconducting channel of {YB\MakeLowercase{a}$_2$C\MakeLowercase{u}$_3$O$_{7-\delta}$} in the mixed state}

\subsection{Electronic structure}

\begin{figure}[b]
\includegraphics[width=\columnwidth]{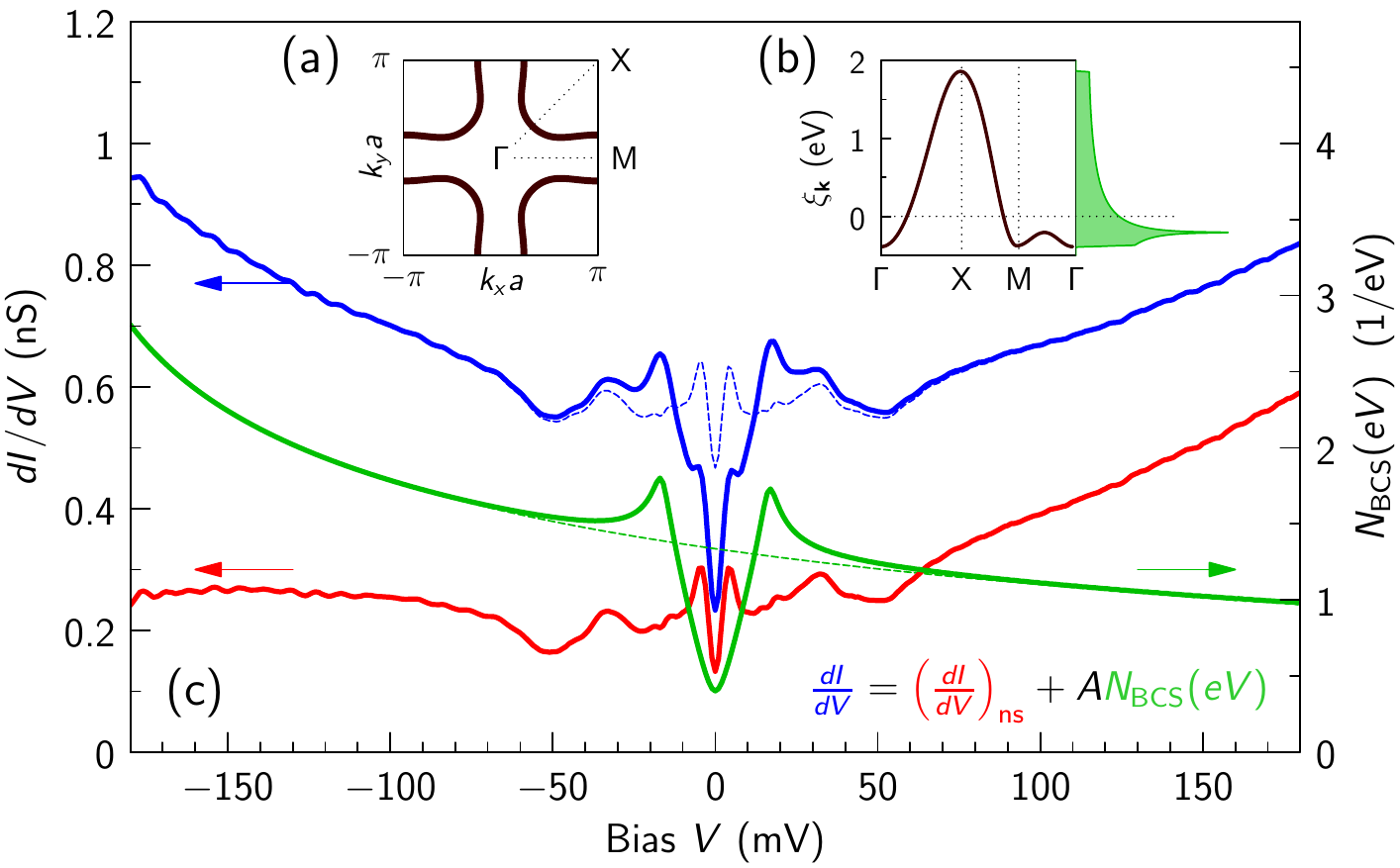}
\caption{\label{fig:uniform}
(a) Fermi surface and (b) dispersion relation. The green curve in (b) is the normal-state DOS with a van Hove singularity at $-376$~meV. (c) Zero-field tunneling conductance of Y123 at 0.4~Kelvin (solid blue, left scale, from Ref.~\onlinecite{Bruer-2016-s}) and its decomposition in superconducting (solid green) and non-superconducting (red) channels. The solid green curve (right scale) is the BCS DOS calculated with the dispersion shown in (b). The finite zero-energy DOS in the gap is due to the finite energy resolution of the calculation ($\approx 3$~meV). The dashed green curve is the corresponding normal-state DOS. The value $A=0.25$~eV~nS is adjusted such that the non-superconducting channel has no coherence peaks at $\pm17$~meV. The dashed blue curve is the sum of the red and dashed green ones.
}
\end{figure}

YBa$_2$Cu$_3$O$_{7-\delta}$ (Y123) has two CuO$_2$ planes in the unit cell, CuO chains running along the $b$ axis, and a small orthorhombic distortion with inequivalent $a$ and $b$ axes ($b>a$) \cite{Jorgensen-1990-s}. We ignore the CuO chains and the distortion, absent in other cuprates, and therefore irrelevant for high-$T_c$ superconductivity. Some effects of the chains on the vortex-core spectra have been studied in Ref.~\onlinecite{Atkinson-2009-s}. We furthermore ignore the bilayer splitting for simplicity and represent the CuO$_2$ layers as a one-band tight-binding model on a perfect square lattice with parameter $a=3.85$~\AA. We have also performed calculations for a two-band system including bilayer splitting, and found only inessential quantitative differences in the vortex cores. The one-band model is more convenient for large-scale simulations. We use the tight-binding parameters $t_1=-281$~meV, $t_2=139$~meV, and $t_3=-44$~meV determined by photoemission in Ref.~\onlinecite{Schabel-1998a-s, *Schabel-1998b-s} for the first, second, and third neighbor hopping, respectively, ignoring $t_4$ and $t_5$ for simplicity. The chemical potential is set to $\mu=-356$~meV for an electron density $n=0.84$, corresponding to optimally hole-doped Y123 with 0.16 hole per unit cell. The dispersion relation measured from the chemical potential is $\xi_{\vec{k}}=2t_1[\cos(k_xa)+\cos(k_ya)]+4t_2\cos(k_xa)\cos(k_ya)+2t_3[\cos(2k_xa)+\cos(2k_ya)]-\mu$. The average group velocity on the Fermi surface is $\langle v_{\mathrm{F}}\rangle=4.11\times10^7$~cm/s. The Fermi surface is shown in Fig.~\ref{fig:uniform}(a). Due to a van Hove singularity at $-376$~meV, the DOS has a negative slope in the low-energy region, with more weight for the occupied states [Fig.~\ref{fig:uniform}(b)]. We set the amplitude of the $d$-wave order parameter to $\Delta_0=19$~meV. In the uniform superconductor, the gap $\Delta_{\vec{k}}=(\Delta_0/2)[\cos(k_xa)-\cos(k_ya)]$ has its maximum at the point $(\pi/a,0.74/a)$ of the Fermi surface, giving coherence peaks at $\pm17$~meV [Fig.~\ref{fig:uniform}(c)]. This amplitude and symmetry of the order parameter follow self-consistently from the Bogoliubov-de Gennes equations for an (instantaneous) attractive interaction $V=-247$~meV on nearest-neighbor bonds.

Figure~\ref{fig:uniform}(c) shows the base-temperature zero-field tunneling spectrum of Y123 \cite{Bruer-2016-s} and a possible decomposition in two channels. The superconducting channel (SC) is calculated with the tight-binding model, and the non-superconducting channel (NSC) is the difference. The relative weight of the two contributions is adjusted in such a way that the NSC has no structure---peak or dip---at the edges of the superconducting gap. The resulting NSC has an overall positive slope. The latter is sensitive to the choice of hopping parameters in the SC, and is therefore not a robust feature of the analysis. This slope is irrelevant for our study of vortex cores focusing on energies $\lesssim50$~meV. In contrast, the dips near $\pm50$~meV, the peaks near $\pm30$~meV, and the subgap peaks near $\pm5$~meV are robust properties of the spectrum measured by STS in regions where superconductivity is suppressed \cite{Bruer-2016-s}. The dashed blue line shows the spectrum expected in such a region where the superconducting gap is closed, assuming that the relative weight of the two channels remains unchanged. This is fully consistent with the spectra measured in non-superconducting regions \cite{Bruer-2016-s}. A more elaborate modeling was introduced in Ref.~\onlinecite{Bruer-2016-s}, that involved bilayer splitting as well as an interaction with the spin fluctuations. The main effect of these additional ingredients is to assign (part of) the dips at $\pm50$~meV to the SC rather than the NSC. As these energies are not our main concern and these sophistications are impractical in view of large-scale vortex calculations, here we disregard them.

\subsection{Isolated vortex, self-consistent solution}

We solve the Bogoliubov-de Gennes equations self-consistently with a single vortex at the origin using the method described in Ref.~\onlinecite{Berthod-2016-s}. The reader is referred to Ref.~\onlinecite{Berthod-2016-s} for all practical details, while here we only give the elements specific to the present calculation. As our Hamiltonian extends up to third neighbors, it spreads the wave function on the lattice with a diamond shape. We therefore consider a finite system with diamond shape and linear size $M$, having $1+2M(1+M)$ lattice sites. We use $M=200$ (80\,401 sites) for calculating the self-consistent order parameter and $M=500$ (501\,001 sites) for calculating the local density of states (LDOS). The order parameter requires a smaller system because the coherence length imposes a spatial cutoff. With $M=200$, a Chebyshev expansion order $N=6000$, and termination using the Jackson kernel \cite{Berthod-2016-s}, the calculation retrieves the exact order parameter within 0.1\%. For the LDOS, the spatial cutoff would be set by the mean free path, which is infinite in our model. With $M=500$ and an expansion order $N=2000$, we reach an energy resolution of roughly $3$~meV without perturbations associated with the system's boundaries.

\begin{figure}[tb]
\includegraphics[width=0.96\columnwidth]{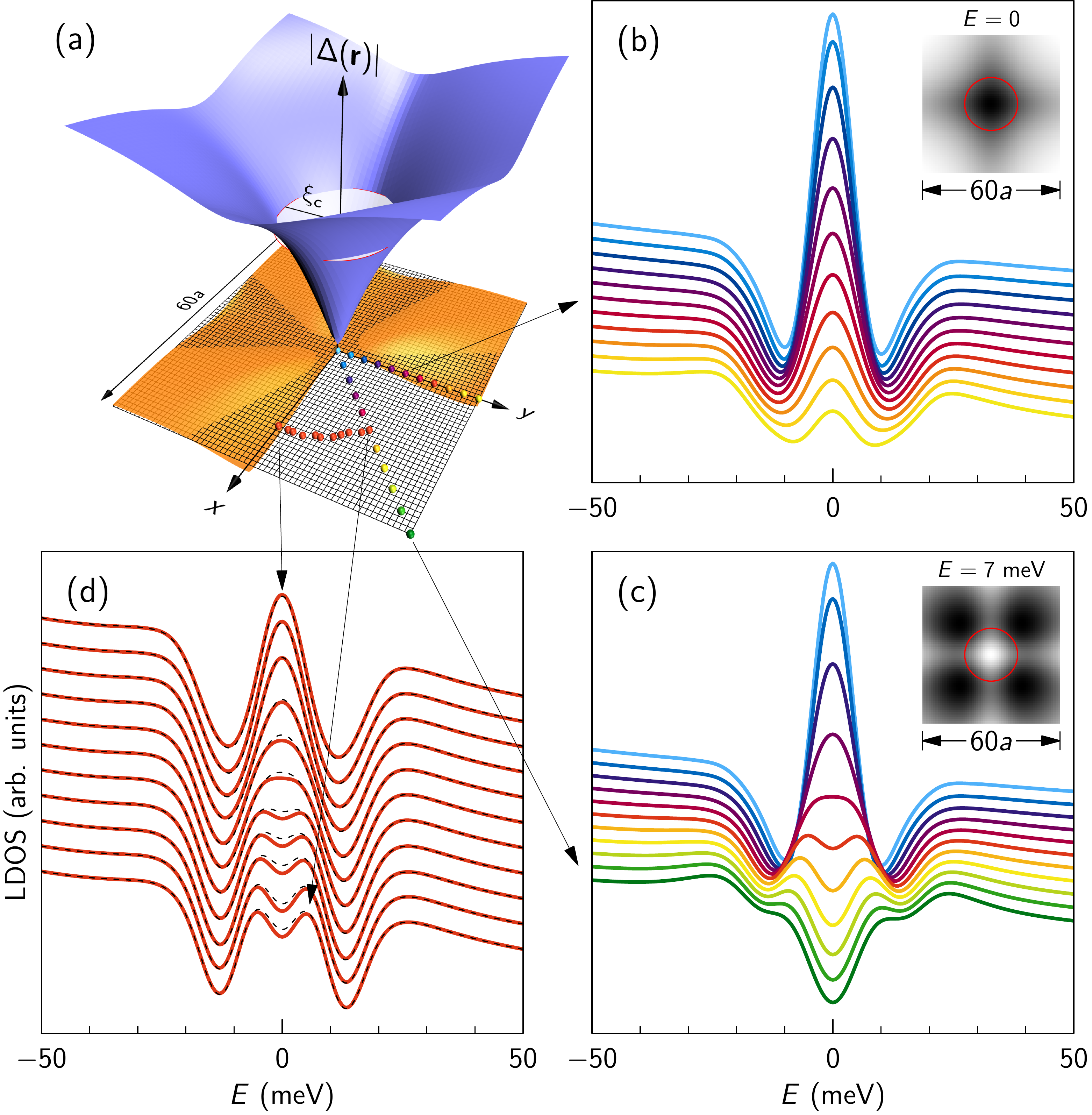}
\caption{\label{fig:vortex}
(a) Modulus of the self-consistent order parameter for an isolated vortex (blue) on each site of the tight-binding lattice (black). The difference between the self-consistent and ansatz solutions (see text) is shown in orange. The white disk indicates the core size, defined as the distance at which the order parameter is $\Delta_0/2$. Colored balls mark the sites where the LDOS is plotted along (b) the antinodal direction, (c) the nodal direction, and (d) at a fixed distance as a function of angle. The LDOS curves are shifted vertically in (b), (c), and (d) and the color encodes the distance to the vortex center. Note that the curves show the full LDOS $N(\vec{r},E)$ without subtraction. The dashed curves in (d) are calculated using the isotropic ansatz for the order parameter. The insets in (b) and (c) show the spatial distribution of the LDOS (low, white to high, black) at the indicated energies, with the red circle of radius $\xi_c$ indicating the core size.
}
\end{figure}

The self-consistent order parameter is plotted in Fig.~\ref{fig:vortex}(a). The quantity $|\Delta(\vec{r})|$ is defined as the sum of the order-parameter modulus on the four bonds surrounding the site $\vec{r}$. It is well fitted by the isotropic ansatz $\Delta(r)=\Delta_0/[1+\xi_0/r\exp(-r/\xi_1)]$ with $\xi_0=17a$ and $\xi_1=29.5a$. The difference between the exact and approximate data is negative along the $x$ and $y$ directions and positive along the diagonals. The self-consistent order parameter indeed displays a small in-plane anisotropy unlike the ansatz, and relaxes faster to its asymptotic value along the diagonals than along the lattice axes, as already found in similar calculations \cite{Ichioka-1996-s, Berthod-2016-s}. A ``core size'' $\xi_c$ may be defined by the condition $\Delta(\xi_c)=\Delta_0/2$, yielding $\xi_c=11.5a=4.4$~nm. This agrees very well with the BCS expression of the coherence length $\xi=\hbar v_{\mathrm{F}}/(\pi\Delta_0)=4.5$~nm if the Fermi-surface average of the velocity is substituted for $v_{\mathrm{F}}$.

The LDOS plotted in Figs.~\ref{fig:vortex}(b), \ref{fig:vortex}(c), and \ref{fig:vortex}(d) displays a zero-energy peak that is maximum at the vortex center and decays differently in all directions. The in-plane anisotropy of the LDOS is not a consequence of the in-plane anisotropy of the order parameter, as illustrated in Fig.~\ref{fig:vortex}(d), where it is seen that the typical angular dependence of the LDOS remains unchanged if an isotropic order parameter is used. It is also not a consequence of the $d$-wave gap symmetry: repeating the calculation for an $s$-wave gap leads to the same anisotropic LDOS with an un-split peak along the antinodal directions and a split peak along the nodal ones. Note that the LDOS peak in Fig.~\ref{fig:vortex} is a genuine continuum, not the superposition of densely packed discrete core levels as in $s$-wave superconductors \cite{Franz-1998b-s, Berthod-2016-s}. In fact, the LDOS anisotropy relates to the dispersion and Fermi-surface anisotropies, which tend to favor low-energy LDOS structures in the directions normal to the Fermi surface. The two traces in Figs.~\ref{fig:vortex}(b) and \ref{fig:vortex}(c) span different distances from the core; Figure~\ref{fig:fig1}(a) of the main text allows one to compare these traces along the same distance. The specific signature of the core for an isolated vortex [Fig.~\ref{fig:fig1}(b) of the main text] is obtained by subtracting the zero-field spectrum (green curve in Fig.~\ref{fig:uniform}) from the vortex LDOS.

\subsection{Ideal vortex lattices, self-consistent solutions}

\begin{figure}[t]
\includegraphics[width=\columnwidth]{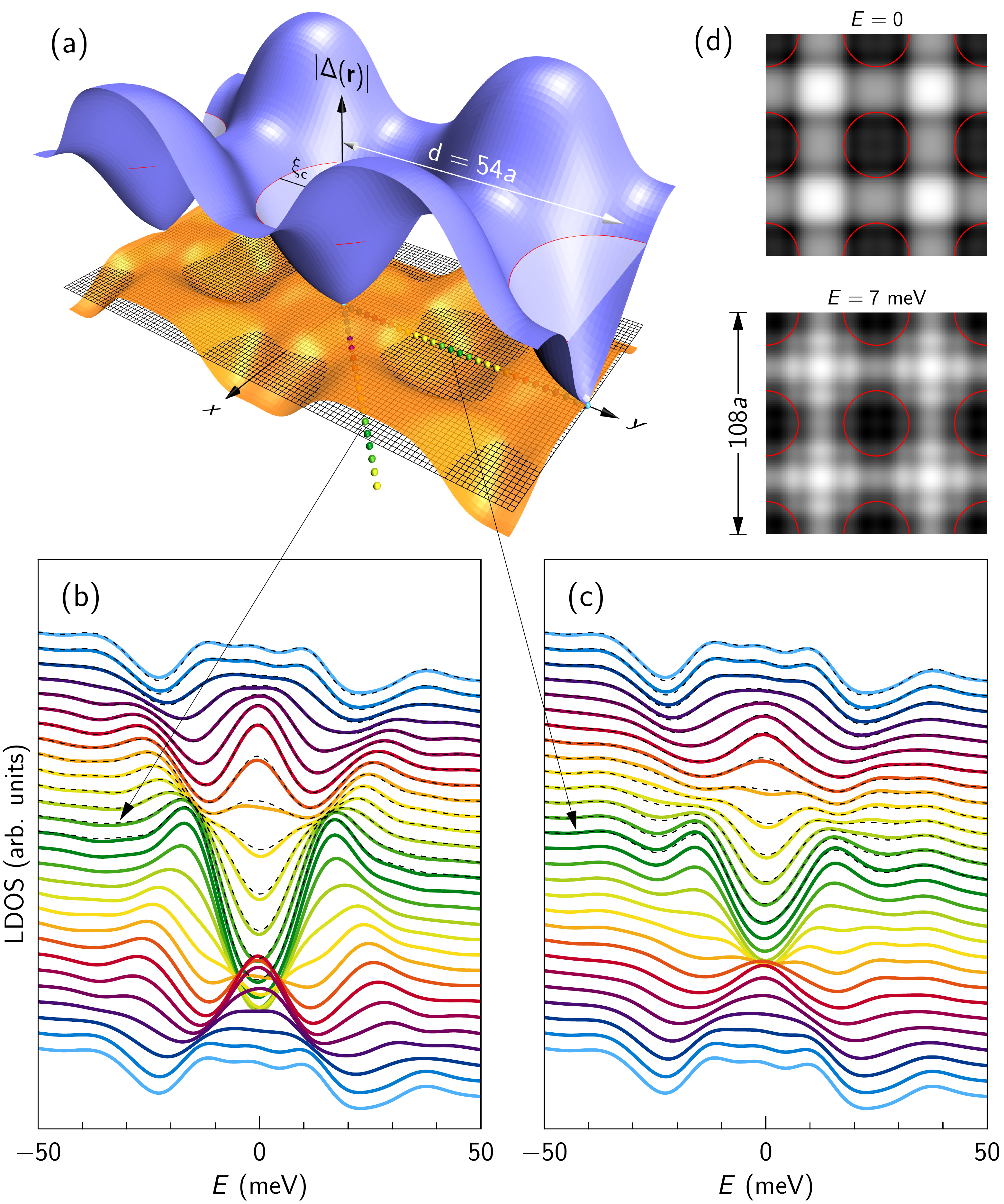}
\caption{\label{fig:lattice-10}
(a) Modulus of the self-consistent order parameter (blue) and difference between the self-consistent and ansatz solutions (orange) for a square vortex lattice with inter-vortex distance $d$ oriented along the principal directions of the microscopic lattice. Two LDOS traces are shown along (b) the line connecting next-nearest neighbor vortices and (c) the line connecting nearest-neighbor vortices. The colors encode the position with respect to the cores as indicated by the balls in (a). The dashed curves, only half of which are shown for clarity, are obtained using the ansatz order parameter instead of the self-consistent one. (d) Spatial distribution of the LDOS at two energies; the red circles of radius $\xi_c$ show the vortex cores.
}
\end{figure}

\begin{figure}[t]
\includegraphics[width=\columnwidth]{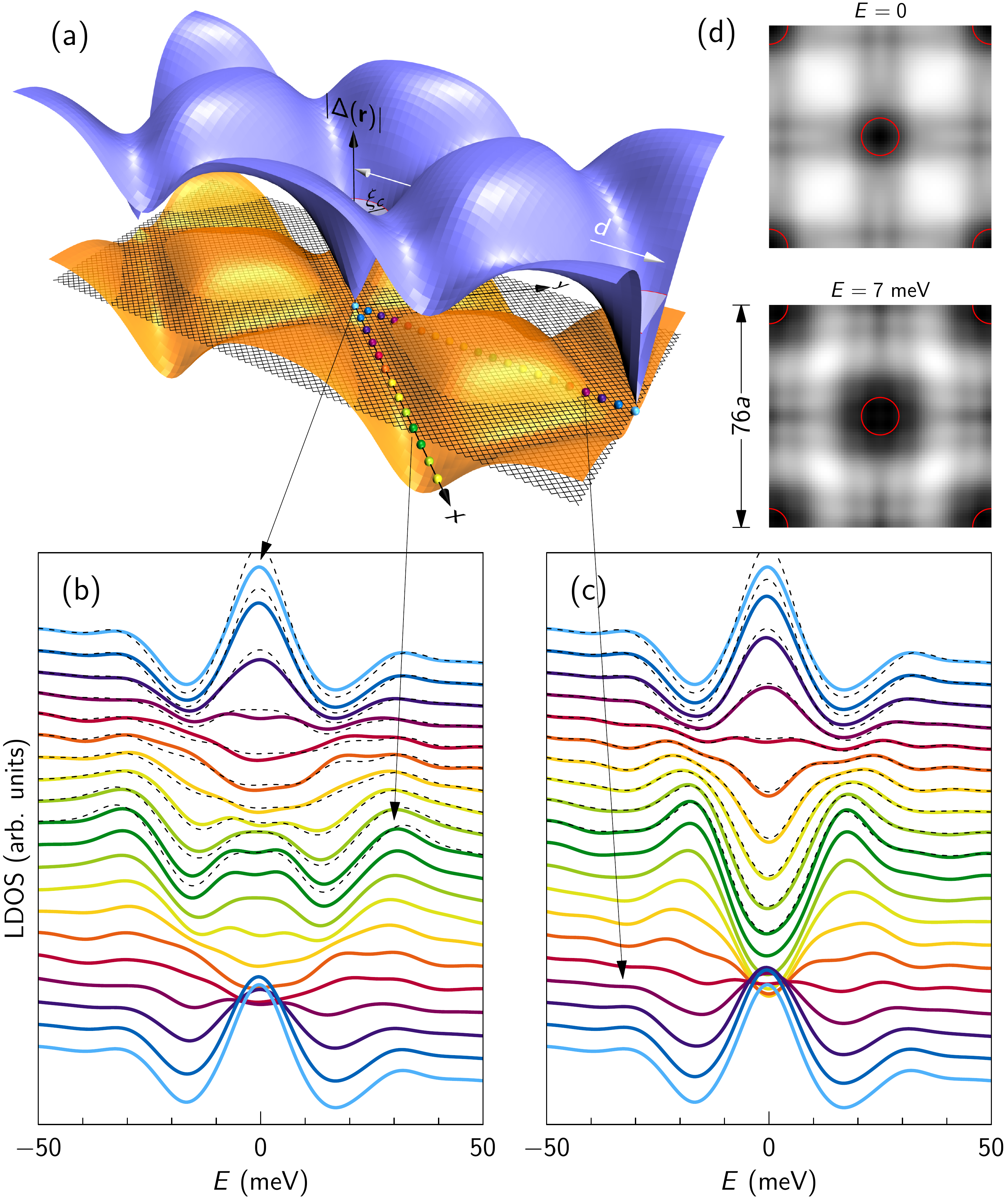}
\caption{\label{fig:lattice-11}
Same as Fig.~\ref{fig:lattice-10} for a square vortex lattice with inter-vortex distance $d=38\sqrt{2}a$ oriented along the diagonals of the microscopic lattice. Note that the microscopic lattice is rotated by 45$^{\circ}$ in the three-dimensional plot (a), as compared to Fig.~\ref{fig:lattice-10}(a). In all LDOS maps of (d) and Fig.~\ref{fig:lattice-10}(d), however, the microscopic lattice directions correspond to the horizontal and vertical directions of the maps.
}
\end{figure}

The self-consistent solution for an ideal square vortex lattice with inter-vortex distance $d=54a$ is displayed in Fig.~\ref{fig:lattice-10}(a). This corresponds to a density of 19 vortices on $90\times 90$~nm$^2$ as observed in the experiment (Fig.~\ref{fig:fig2} of the main text). The vortex lattice is aligned with the microscopic lattice, the nearest-neighbor vortices being found along the $x$ and $y$ directions. The order parameter is maximum in the center of the squares formed by four nearest-neighbor vortices and has saddle points with $|\Delta(\vec{r})|=12$~meV on the lines joining them, leading to a significant spatial modulation. Note that the solution has been constrained to have a maximum gap of 19~meV like in zero field for simplicity---and also because no measurable reduction of the gap size is observed experimentally at this field---requiring a slight increase of the interaction to $V=-260$~meV. $|\Delta(\vec{r})|$ has a rounded shape in the cores, which is captured by the ansatz (generalized for vortex lattices \cite{Berthod-2016-s}) with an increased value $\xi_0=115a$ and a reduced value $\xi_1=8a$ with respect to the isolated vortex. The core size defined as $\Delta(\xi_c)=\Delta_0/2$ is slightly increased to $\xi_c=15.9a=6$~nm compared with the isolated vortex, consistently with previous studies in the quantum regime \cite{Berthod-2016-s}.

The corresponding LDOS is shown in Figs.~\ref{fig:lattice-10}(b), \ref{fig:lattice-10}(c), and \ref{fig:lattice-10}(d). Although the relative difference between the exact and ansatz solutions is 12\% at maximum, the LDOS curves calculated with both order parameters are almost undistinguishable. This is the justification for using the non-self-consistent ansatz when studying the LDOS in disordered vortex configurations, for which a full self-consistent calculation is impractical. The zero-energy LDOS peak is considerably suppressed and broadened in the core with respect to the isolated vortex. We have checked that the vortex-lattice calculation correctly reproduces the isolated-vortex limit as the distance $d$ is increased: both spectra differ by less than 5\% for $d\gtrsim 170a$ ($B\lesssim 0.5$~T). At the field considered, however, both the spectral and spatial signatures of the vortex cores differ markedly from those of the isolated vortex seen in Fig.~\ref{fig:vortex}. We have also verified that the suppression of the zero-energy LDOS peak is not due to the increased value of $\xi_c$ and more rounded order parameter in the core: the spectra shown in Fig.~\ref{fig:lattice-10} remain qualitatively unchanged if we use the ansatz order parameter with the values of $\xi_0$ and $\xi_1$ corresponding to the isolated vortex. The reason for a suppressed vortex-core peak can be understood by comparing the low-energy LDOS in Figs.~\ref{fig:vortex}(b) and \ref{fig:lattice-10}(d). Because for the isolated vortex the LDOS extends farther along the (10) direction than along the (11) direction, in the vortex lattice the wavefunctions in different cores strongly overlap and the core states get delocalized. This overlap is suppressed when the vortex lattice is not precisely aligned with the (10) direction and/or the vortex positions are disordered, such that the zero-energy LDOS peak is restored in these situations (see below).

For comparison, we show in Fig.~\ref{fig:lattice-11} the self-consistent order parameter and LDOS for a square vortex lattice rotated 45$^{\circ}$ with respect to the tight-binding lattice with an inter-vortex distance $d=38\sqrt{2}a$, which corresponds to the same field as in Fig.~\ref{fig:lattice-10}. There are significant differences between the two vortex-lattice orientations (hereafter I and II), both in the self-consistent order parameter and in the LDOS. While in I the gap has saddle points between nearest-neighbor vortices and maxima between next-nearest-neighbor ones, the situation is reversed in II with the gap maxima between nearest-neighbor vortices. It appears that the order parameter doesn't move rigidly with the vortex lattice: when rotating the vortex-lattice orientation from I to II, the cores move but the gap maxima and saddle points stay in place. The shape of the core is also quite different in I and II, where a best fit to the ansatz gives $\xi_0=9.9a$, $\xi_1=15a$, and a core size $\xi_c=6.4a=2.5$~nm smaller than in zero field. There is more structure in the case II, because each saddle point is in fact replaced by two saddle points separated by a local minimum. This explains the larger discrepancy between the ansatz and the exact solution, which reaches 32\% for II at the local minima. As a result, the differences between the LDOS calculated with the ansatz and self-consistent order parameters are slightly larger in II than in I. These differences remain nevertheless small compared with the qualitative differences between the LDOS in I and II: the zero-energy peak is neither strongly suppressed nor split in II as it is in I; at the position of the local minimum between two saddle points in II, the LDOS has a double peak at zero energy, while at the saddle point in I the LDOS is gapped. We see in Fig.~\ref{fig:lattice-11}(d) that the low-energy LDOS is much more localized in the cores compared with I, which highlights the much weaker wavefunction hybridization along the (10) directions in case II.

The data in Figs.~\ref{fig:vortex}, \ref{fig:lattice-10}, and \ref{fig:lattice-11} demonstrate that the vortex-core LDOS is not only a function of field, but also and more importantly a function of the positions of neighboring vortices. Depending on where the neighbors are, the zero-energy LDOS peak may be sharp or not, split or not, etc. Experimentally, one therefore expects variability in the measured vortex-core spectra when the vortex positions are disordered. These figures also show that, in order to investigate theoretically this variability, it is sufficient to work with the ansatz order parameter, whose only inputs are the vortex positions and the values of $\xi_0$ and $\xi_1$.

\section{\boldmath LDOS calculations for disordered vortex lattices }

\subsection{Disordered vortex configurations}

\begin{figure}[b]
\includegraphics[width=\columnwidth]{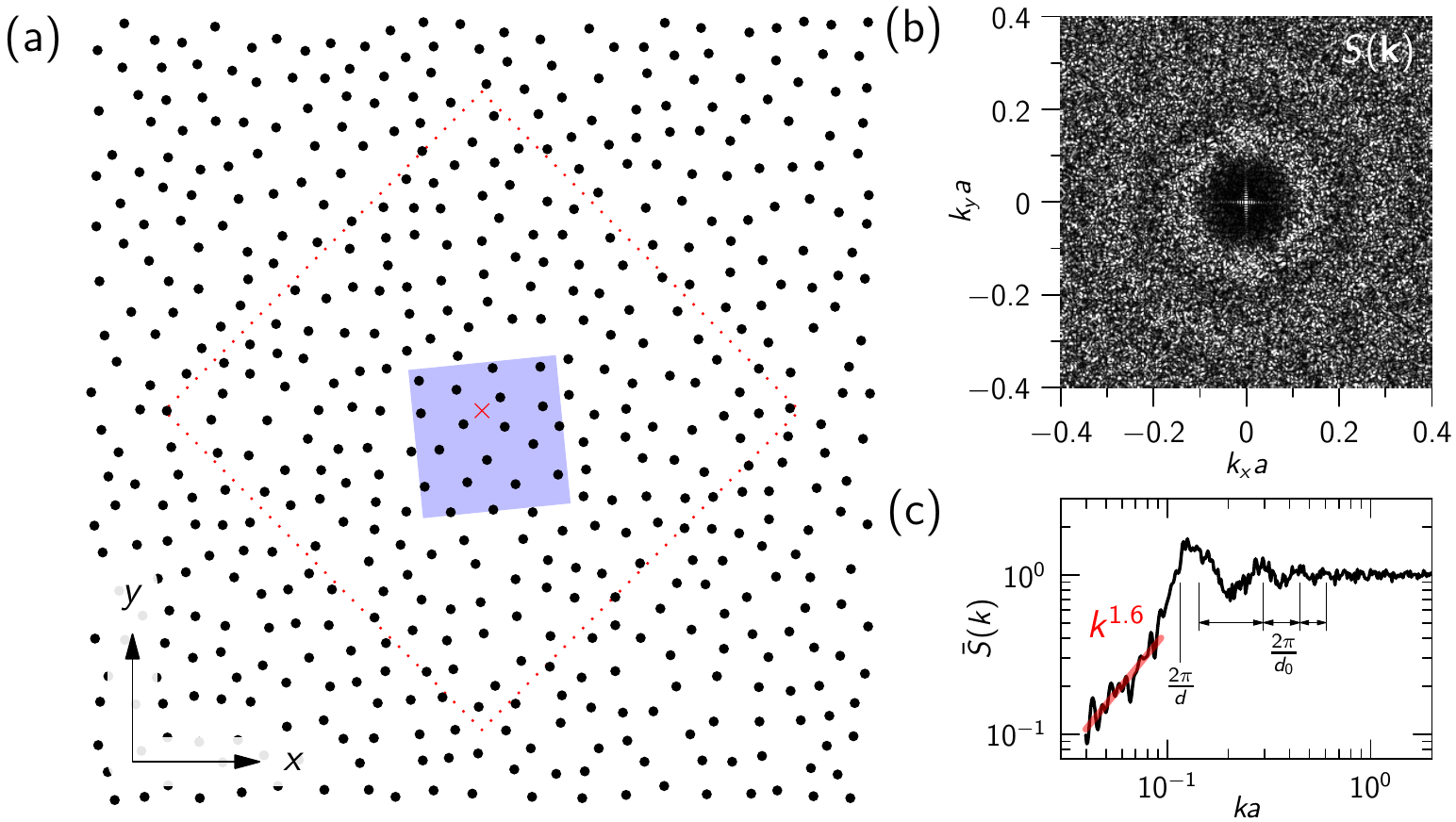}
\caption{\label{fig:disorder}
(a) Typical disordered vortex configuration. The central square represents the STS field of view of $90\times90~\mathrm{nm}^2$, where the vortices are located as observed in the experiment. The square is rotated by 5.7$^{\circ}$ with respect to the crystal axes. The dotted red square indicates the system size used for calculating the LDOS at the point marked by a cross; as the cross moves, the dotted square moves with it. (b) Isotropic structure factor showing the absence of orientational order in the generated vortex positions. (c) Angular average of the structure factor. The power-law behavior for $k<2\pi/d$ is shown in red.
}
\end{figure}

\begin{figure*}[tb]
\includegraphics[width=0.7\textwidth]{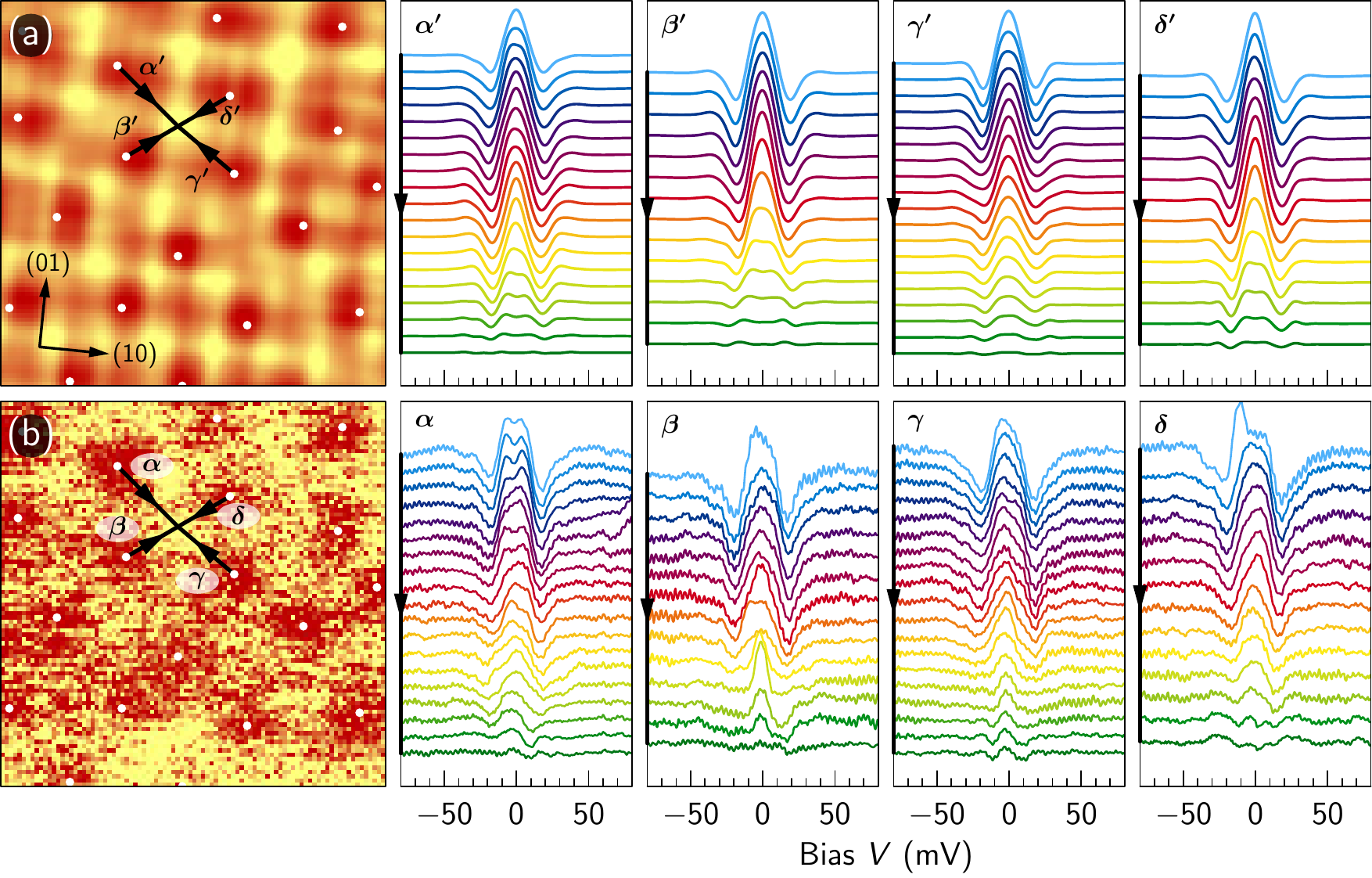}
\hfill\parbox[b]{0.25\textwidth}{
\caption{\label{fig:optimization}
For each configuration of the vortices outside the STS field of view, the theoretical spectral traces calculated along the paths $\alpha'$--$\delta'$ shown in (a) are compared with the corresponding experimental traces along $\alpha$--$\delta$ shown in (b).
}
\vspace{5.4cm}}
\end{figure*}

The high sensitivity of the theoretical LDOS to vortex positions prompts us, for a meaningful comparison with the STS experiment, to use in the calculation the vortex positions as they appear under the STM. Three difficulties arise: (i) the LDOS maxima which are accessible experimentally may not sit exactly on the phase singularity points where the order parameter vanishes, due to a polarization of the LDOS by asymmetric supercurrents \cite{Berthod-2013a-s, Berthod-2016-s}; (ii) we cannot disregard vortices that are outside the STS field of view although we don't know their positions; and (iii) due to lack of atomic resolution on the Y123 surface, the orientation of the microscopic lattice is only known approximately via the macroscopic twin boundaries. We ignore (i), expected to be a small effect, and locate the vortices inside the STS field of view at the positions of largest $dI/dV$ contrast [see Fig.~\ref{fig:fig4}(a) of the main text]. Outside the field of view, we generate vortex positions with the same density as inside, which corresponds to a field of 4.85~T. The positions are chosen at random, however, with a hard-core repulsion constraining the inter-vortex distance to be at least $d_0=41a$. This value was selected to be as large as possible: for larger values the random generation process would be stuck, not able to fit in the required number of vortices. The resulting vortex configurations show some degree of order, similar to what is seen in the STS field of view. An example is shown in Fig.~\ref{fig:disorder}. The structure factor $S(\vec{k})=\mathscr{N}^{-1}\left|\sum_ne^{-i\vec{k}\cdot\vec{R}_n}\right|^2$, where $\vec{R}_n$ are the positions of the $\mathscr{N}$ vortices, is isotropic indicating no orientational order. The angular average $\bar{S}(k)=\mathscr{N}^{-1}\sum_{nm}J_0(k|\vec{R}_n-\vec{R}_m|)$, where $J_0$ is the Bessel function, shows oscillations of wavevector $2\pi/d_0$ due to the hard-core repulsion. Furthermore, a power-law suppression of $\bar{S}(k)$ is observed for $k<2\pi/d$, where $d=54a$ is the inter-vortex distance in the equivalent ordered square lattice. Such power law is reminiscent of hyperuniformity, i.e., a type of order characterized by the suppression of density fluctuations at long wavelengths \cite{Torquato-2003-s}, where $\bar{S}(k)\sim k^{2-\eta}$ with $0<\eta\leqslant 2$ in two dimensions. It is likely that in reality the vortices outside the field of view present more order than the configurations generated by our procedure \cite{Maggio-Aprile-1995-s}, but the experimentally available vortex positions are not sufficient for inferring such an order. While certain characteristics of the vortex ordering outside the field of view may influence the LDOS inside, we do not expect this to change any of the conclusions we draw from our analysis.

\subsection{Search for a good configuration of vortices}

\begin{figure*}[tb]
\includegraphics[width=0.7\textwidth]{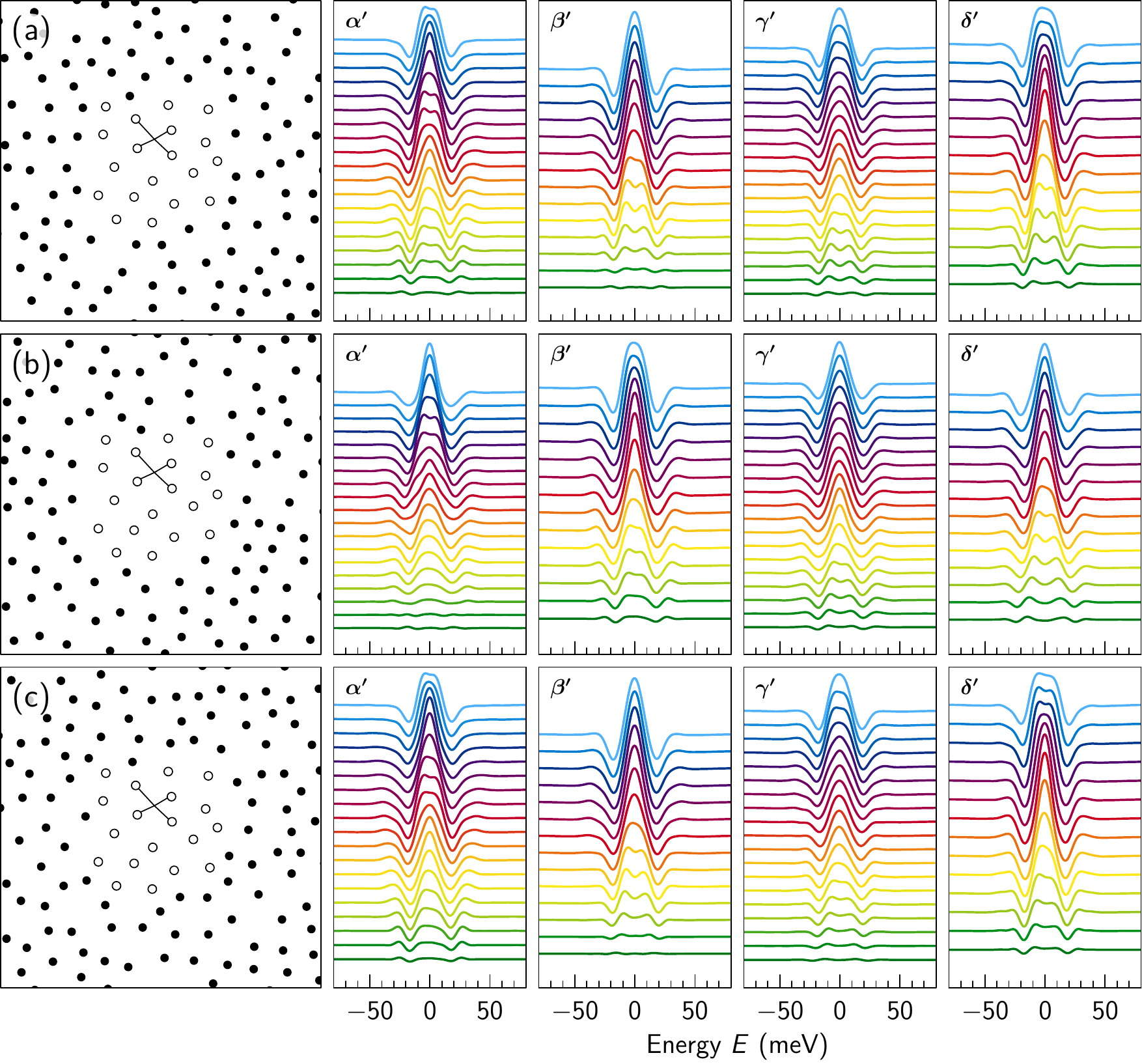}
\hfill\parbox[b]{0.25\textwidth}{
\caption{\label{fig:statistics}
(a), (b), and (c) display three distributions of the vortices outside the STS field of view (black dots), and the vortices at fixed positions within the field of view (white dots). The corresponding spectral traces along the paths $\alpha'$--$\delta'$ of Fig.~\ref{fig:optimization} are plotted on the right.
}
\vspace{8.35cm}}
\end{figure*}

\begin{figure*}[tb]
\includegraphics[width=1\textwidth]{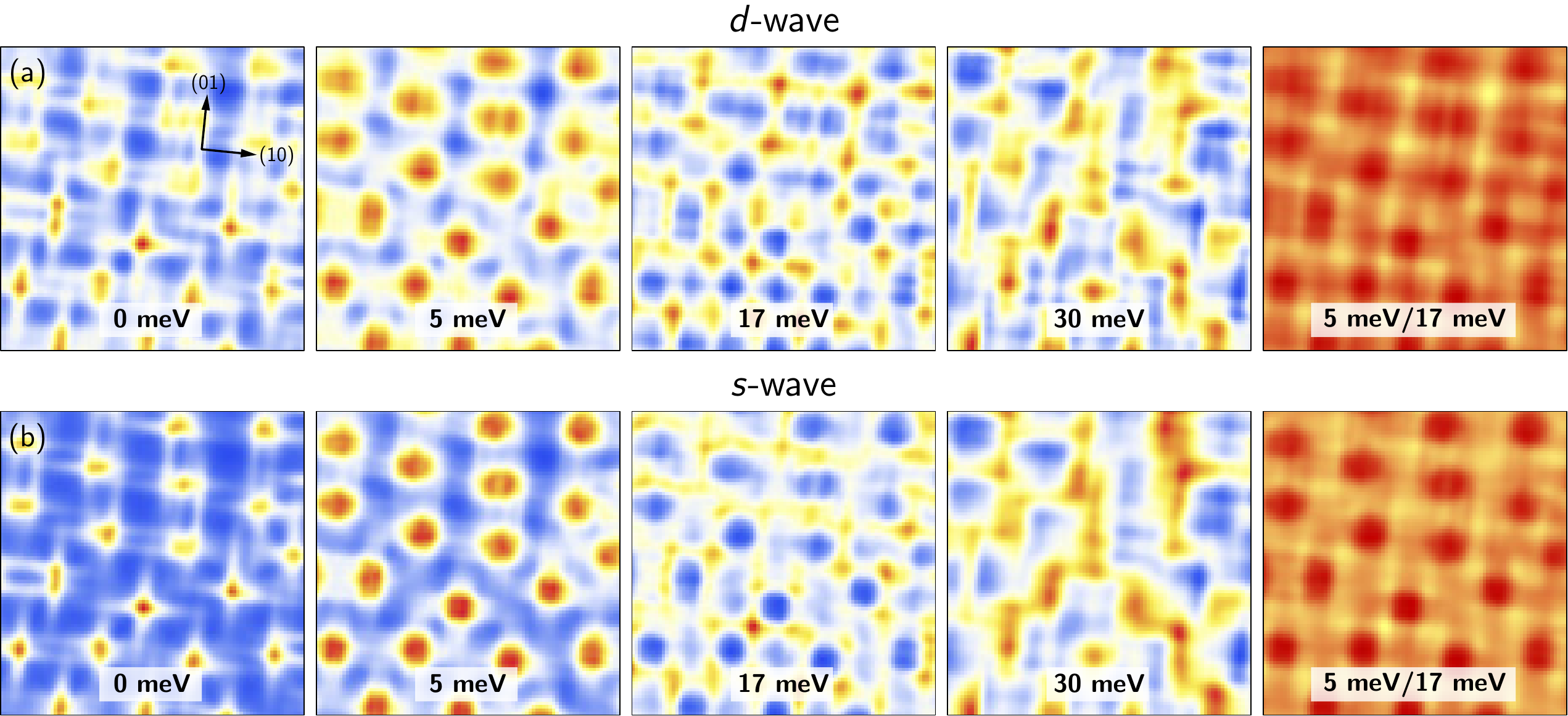}
\caption{\label{fig:swave}
LDOS calculated in the configuration of Fig.~\ref{fig:disorder}(a) for a superconducting gap of (a) $d_{x^2-y^2}$ and (b) $s$ symmetry. Apart from the gap symmetry, all model parameters are identical in (a) and (b). The LDOS is shown at various energies in different panels. The rightmost panels show the ratio as calculated in Fig.~\ref{fig:optimization}(a), but with the colormap extending from the minimum to the maximum of the data, hence a slightly different contrast.
}
\end{figure*}

We have generated 600 disordered vortex configurations for various orientations of the microscopic lattice with respect to the STS field of view. For each configuration, we generate the order parameter using the ansatz and the values of $\xi_0$ and $\xi_1$ corresponding to Fig.~\ref{fig:lattice-10}, and we calculate the LDOS along the four paths displayed in Fig.~\ref{fig:optimization}(a), that correspond to paths n\textsuperscript{o}6 and n\textsuperscript{o}8 in Fig.~\ref{fig:fig4} of the main text. These four traces share a common point [the cross in Fig.~\ref{fig:disorder}(a)]: we use the LDOS at this point as the reference spectrum and subtract it from the calculated LDOS along the four paths. The same procedure applied to the experimental data [Fig.~\ref{fig:optimization}(b)], allows us to compute a sum of squared differences as the figure of merit for each vortex configuration. We find that the agreement between measurement and simulation is systematically better when the four paths are close to a nodal direction. The best compromise is reached if the STS field of view is rotated by 5.7$^{\circ}$ relative to the microscopic lattice as shown in Fig.~\ref{fig:disorder}(a). This points to a tendency for the nearest-neighbor vortices to align along the crystal axes, and justifies a posteriori our use of the values of $\xi_0$ and $\xi_1$ corresponding to that orientation rather than that of Fig.~\ref{fig:lattice-11}. We emphasize again that the precise values of $\xi_0$ and $\xi_1$ play very little role in the LDOS. Figure~\ref{fig:disorder}(a) is the best among the 600 configurations; the four traces are compared with the experimental ones in Fig.~\ref{fig:optimization}. Using this configuration, we calculate the LDOS in the whole STS field of view with 1~nm resolution. For comparing with Fig.~\ref{fig:fig4}(a) of the main text, we compute $dI/dV$ by adding to our theoretical LDOS the non-superconducting channel using the formula quoted in Fig.~\ref{fig:uniform}, we evaluate the ratio between the calculated $dI/dV$ at 5 and 17~meV, and thus obtain the map and traces shown in Fig.~\ref{fig:optimization}(a) and Fig.~\ref{fig:fig4}(b) of the main text.

Figure~\ref{fig:statistics} displays three vortex configurations different from the best one shown in Fig.~\ref{fig:disorder}(a) and \ref{fig:optimization}, and the corresponding theoretical spectral traces. These configurations agree reasonably with experiment as well, with a figure of merit within the best 10\% out of the 600 considered. One notices, in particular, four spectroscopically very different cores with configuration (a), a reinforcement of the LDOS at intermediate distance in (b), trace $\alpha'$, similar to what is seen in Fig.~\ref{fig:fig2}(c) of the main text, and split spectra at the center of vortices $\alpha'$ and $\delta'$ for configuration (c), as seen in the experiment [Fig.~\ref{fig:optimization}(b)].

\subsection{Mixed-state LDOS and gap symmetry}

It is tempting to search signatures of the $d_{x^2-y^2}$ symmetry of the order parameter in the LDOS around vortices. In the semiclassical regime $k_{\mathrm{F}}\xi\gg1$, the vortices are slowly varying perturbations of the order parameter compared with atomic distances, and their Fourier components are mostly at low momenta. Therefore, the vortices provide only small momentum transfers and the interaction of the Bogoliubov quasiparticles with the vortex lattice is dominated by forward scattering. In that limit, the details of the Fermi surface are irrelevant and the only source of spatial LDOS anisotropy---apart from the vortex lattice itself---is indeed the order-parameter symmetry. In the quantum regime $k_{\mathrm{F}}\xi\sim1$ relevant for Y123, however, the vortex lattice scatters Bogoliubov quasiparticles with large momentum transfers of the order of $k_{\mathrm{F}}$ and the LDOS is therefore sensitive to the anisotropy of the Fermi surface. This situation generically leads to the development of LDOS structures along the directions normal to the Fermi surface, a fact well known from studies of impurity scattering \cite{Weismann-2009-s}. In Y123, the Fermi surface segments are mainly oriented along the crystallographic directions [see Fig.~\ref{fig:uniform}(a)], such that one expects structure in the vortex-lattice LDOS along the (10) and (01) lattice directions, irrespective of the order-parameter symmetry. This is confirmed by our numerical results shown in Fig.~\ref{fig:swave}: for both $d$- and $s$-wave symmetries, the LDOS structures are aligned with the microscopic lattice at all energies. No structure is observed along the (11) and equivalent directions, which are the directions of the gap nodes in reciprocal space. Thus the expectation that the LDOS would ``leak'' out of the vortices along the directions of the gap nodes is not confirmed. There are nevertheless differences between the LDOS calculated for $d_{x^2-y^2}$ and $s$ symmetries. The vortex states are more localized in the $s$-wave case, leading to a better contrast, especially at low energy. However, according to these simulations, an unambiguous determination of the order-parameter symmetry based on experimental LDOS maps around vortices appears to be hopeless.

\end{document}